\input{aipcheck}

\documentclass[
    ,final            
  ]
  {aipproc}

\layoutstyle{6x9}

\begin{document}

\title{Shortcuts in extra dimensions and neutrino physics}


\author{Heinrich P\"as}{
  address={Department of Physics \& Astronomy,
University of Alabama,
Tuscaloosa, AL 35487,
USA 
}}

\author{Sandip Pakvasa}{
  address={Department of Physics \& Astronomy, 
University of Hawaii at Manoa,
2505 Correa Road, Honolulu, HI 96822, USA
}}

\author{Thomas J. Weiler}{
  address={Department of Physics \& Astronomy, 
Vanderbilt University, Nashville, TN 37235, USA}
}

\begin{abstract}
We discuss the possibility of sterile neutrinos taking shortcuts in extra 
dimensions, and the consequences for neutrino oscillations. This effect
influences the active-sterile neutrino mixing and 
provides a possibility to accommodate the LSND evidence for neutrino 
oscillations together with bounds from accelerator and reactor experiments.
We briefly comment on causality in such schemes.
\end{abstract}

\maketitle

A typical feature of theories with large extra dimensions is the idea to avoid
the large gauge hierarchy by lowering the Planck scale. Such absence of any
large energy scale in the theory invalidates the most prominent
seesaw suppression of neutrino masses and requires an alternative
mechanism for neutrino mass generation.
However, string theories usually predict singlet fermions in the bulk
(e.g. superpartners of moduli fields) which have the right quantum numbers
to act as right-handed neutrinos. This picture leads to a suppressed wave 
function overlap between the left-handed neutrino localized on a 3+1
dimensional brane and its right-handed partner, and
small neutrino masses can be 
obtained from volume-suppressed  couplings to right-handed neutrinos 
in the bulk \cite{add}.

Another interesting feature of extra-dimensional theories has been first 
proposed as an alternative to inflationary cosmology. Such an era of 
exponential expansion in the early universe  
explains why the universe seems to be homogenous over distances without causal contact,
and is basic to the concordance model.
The alternative idea put forward in \cite{shortcuts} 
is that bulk gravitons may take
shortcuts in the extra dimensions and this way establish causal contact
between space-time regions apparently causally separated for paths
traversing the brane.

One can easily sketch three different mechanisms for such bulk shortcuts:

\begin{itemize}

\item
Self-gravity effects in the presence of matter localized on the brane
imply extrinsic brane curvature.

\item Thermal or quantum fluctuations lead to brane bending.

\item 
The extra dimension can be warped asymmetrically, i.e. warp factors can shrink 
the space dimensions $x$ parallel to the brane but leave the time and bulk 
dimension $t$ and $u$
unaffected, as in the following example \cite{shortcuts,aswarp}:
\begin{equation}
ds^2 = dt^2 -[e^{-2 k u}a^2(t) d x^2 + du^2].
\label{warpmet}
\end{equation}

\end{itemize}

In a recent paper we have analyzed the effect of such shortcuts 
on neutrino oscillations in the presence of sterile bulk neutrinos, and
its consequences for the LSND dilemma
\cite{Pas:2005rb}.
The LSND experiment has reported evidence for $\overline{\nu}_\mu
\rightarrow \overline{\nu}_e$ oscillations with 
$\delta m^2_{\rm LSND}\sim 1$~eV$^2$.
In order to accommodate the LSND result with the confirmed solar 
($\delta m^2_{\rm sun}\sim 10^{-4}$~eV$^2$) and atmospheric 
($\delta m^2_{\rm atm}\sim ({\rm few})10^{-3}$~eV$^2$) neutrino
results, a 4-th neutrino is 
required. Moreover, since the width of the Z-boson as measured by LEP allows 
only for three neutrinos coupling to the Z, this 4-th neutrino 
has to be  
sterile. Thus the four states can be arranged as
either the two pairs of solar and atmospheric 
neutrinos, separated by the large LSND mass gap (the 2+2 spectrum)
or as a triplet of active neutrinos complemented by a single, predominantly 
sterile state (the 3+1 spectrum).
Hereby the first option is excluded, since neither solar nor 
atmospheric neutrinos have been observed to oscillate into steriles.
On the other hand, the second option is strongly constrained by the negative
results of reactor (BUGEY) and accelerator (CDHS) experiments searching for 
electron or muon neutrino disappearance. 

The oscillation amplitudes for BUGEY, CDHS and LSND are approximately given by
$\sin^2 2 \theta_{e\not{e}}\approx 
4\,U_{e 4}^2$, 
$\sin^2 2 \theta_{\mu\not{\mu}}\approx 4\,U_{\mu 4}^2$,
and 
$
\sin^2 2 \theta_{\rm LSND} = 4 U_{e4}^2 U_{\mu 4}^2$, where the mixing 
matrix elements $U_{e/\mu 4}$ parametrize the admixture of $\nu_e$, $\nu_\mu$ 
in the 4-th state.
Consequently the LSND amplitude is doubly suppressed by the 
bounds obtained from
BUGEY and CDHS, invalidating the 3+1 spectrum.
While one 
solution  to the LSND dilemma is that LSND
might be wrong (results of the ongoing test at MinibooNE are due!),
it also
may hint towards deviations from the usual oscillation mechanism
 and
may be a messenger of the mechanism of neutrino mass generation.
A particular attractive example for the latter case are
extra dimensions and bulk shortcuts.

If we consider neutrino oscillations in the presence of bulk shortcuts
the evolution factor $\int H dt$ in the path integral is altered, as the
bulk signal gains a time shift $\Delta t$, implying
\begin{equation}
\Delta \int H dt = H \Delta t \rightarrow \Delta H_{\rm eff} T.
\end{equation}
By introducing the shortcut parameter 
$\epsilon \equiv (t_{\rm brane}- t_{\rm bulk})/ t_{\rm brane} 
= \Delta t/T$ the effective Hamiltonian in flavor space is given by
\begin{equation}
\Delta H_{\rm eff}=
+ \frac{\delta m^2}{4\,E}\, \left(
\begin{array}{cc}
 \cos 2\theta  & -\sin 2\theta \\
-\sin 2\theta  & -\cos 2\theta \\
\end{array}
\right)
+E\,\frac{\epsilon}{2}\,\left(
\begin{array}{cc}
-1 & 0 \\
 0 & 1 \\
\end{array}
\right).
\end{equation}

It is obvious, that for a certain resonance energy the shortcut term cancels 
the diagonal piece of the brane Hamiltonian, implying a 
resonance at
$E_{\rm res} = \sqrt{\frac{\delta m^2\,\cos 2\theta}{2\,\epsilon}}$.
Then the
oscillation amplitude becomes
\begin{equation}
\sin^2 2\tilde{\theta} =  
\left[\frac{\sin^2 2\theta}{\sin^2 2\theta +\cos^2 2\theta(1-A)^2}\right]
\end{equation}
with
$A=(E/E_{\rm res})^2$. This implies that 
oscillations at energies $E\gg E_{\rm res}$ are suppressed, and that
the CDHS bound with $E_{\rm CDHS} \gg E_{\rm LSND}$ may not apply to
oscillations as observed at LSND. Thus, in the presence of bulk shortcuts, the
3+1 spectrum is allowed again.

Results for the LSND oscillation probability for two limiting cases 
($E_{\rm res}=33$~MeV and $E_{\rm res}=400$~MeV) are shown in 
Figs.~1 and 2. The LSND energy spectrum in both cases 
resembles the case of standard 
oscillations, and the constraint from the KARMEN experiment reporting a 
null result, is satisfied.
The large resonance energy of the latter case leads to strongly enhanced muon 
neutrino oscillations at MiniBooNE as shown in Fig.~3.
The small resonance of the first
case predicts no neutrino oscillations at MiniBooNE, but does predict 
a distorted LSND spectrum and enhanced oscillations at a proposed experiment at 
the SNS \cite{sns}.

\begin{figure}
  \includegraphics[height=.25\textheight]{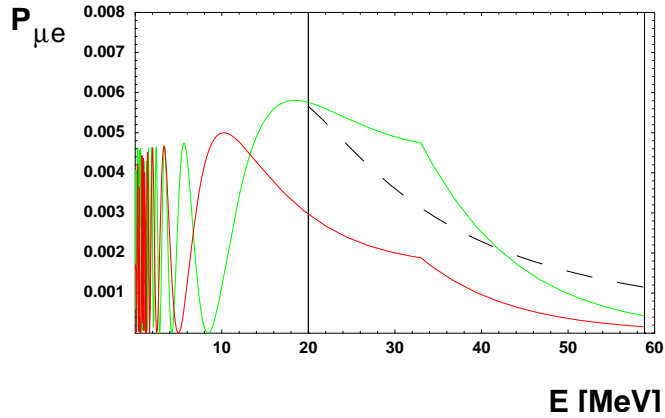}
  \caption{Bulk shortcut oscillation probabilities for LSND (light/green) and 
KARMEN (dark/red, reporting a null result) as a function of the 
neutrino energy, for
$E_{\rm res}=33$~MeV; For comparison, a standard oscillation probability for 
LSND is displayed (dashed).}
\label{eres33p}
\end{figure}

\begin{figure}
  \includegraphics[height=.25\textheight]{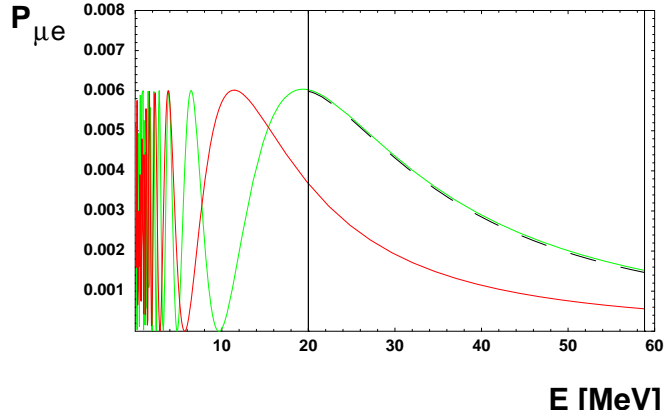}
  \caption{As Fig.~1, but for $E_{\rm res}=400$~MeV.}
\label{eres400p}
\end{figure}

\begin{figure}
  \includegraphics[height=.25\textheight]{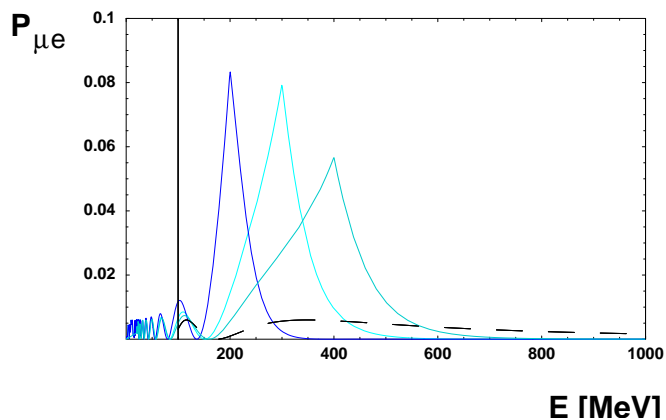}
  \caption{
Bulk shortcut oscillation probabilities for MiniBooNE 
as a function of the neutrino energy, for
$E_{\rm res}=200,~300,~400$~MeV,
from left to right (dark to light).
For comparison, the expectation for a standard oscillation solution 
for LSND is displayed (dashed). 
}
\label{minip}
\end{figure}


In conclusion,
we have discussed the effect of bulk shortcuts, which arise quite naturally in
extra-dimensional theories, on neutrino oscillations. In this framework
bulk shortcuts affect neutrino mixing and imply a new resonance, which implies
a suppression of the oscillation amplitude at energies $E\gg E_{\rm res}$.
This effect accommodates the LSND result in a consistent 
framework with
results from other neutrino oscillation experiments, and predicts strongly 
enhanced oscillations at MiniBooNE for $E_{\rm res} \gg 100$~MeV.
Moreover the BBN bound on a sterile neutrino abundance in the early universe
may be evaded, if bulk shortcut effects like gravitational brane bending,
brane fluctuations, or scattering off our brane into an asymmetrically warped
 bulk are
amplified by higher temperatures and densities in the early universe 
in a manner which lowers
the resonance energy below the MeV scale. In this case, oscillations into 
sterile neutrinos would be suppressed at BBN, the sterile state would not be 
populated, and fatal cosmological consequences would be evaded.

Open questions concern the details of BBN, supernova neutrinos, effects on
atmospheric neutrino oscillations, realistic 3+1 neutrino fits, the relation
of the LSND neutrino to cosmological dark matter, and influences on the
horizon problem. We have started to analyze causality properties of
shortcut spacetimes, and have pointed out that certain examples of asymmetrically 
warped spacetimes may allow for closed timelike curves 
\cite{Pas:2006si}. 
It is an attractive
thought that sterile neutrinos, if their existence is confirmed in 
future experiments,
may help to experimentally probe Hawking's chronology protection conjecture,
or even to realize neutrino time travel.

\bibliographystyle{aipprocl}

\end{document}